\documentstyle[11pt,emulateapj]{article}
\submitted{Accepted for Publication in {\em The Astrophysical Journal}}

\received{}
\accepted{}

\newcommand{\lya}		{Ly$\alpha$}
\newcommand{\lyb}		{Ly$\beta$}
\newcommand{\qso}		{{Q0122+0338}}
\newcommand{\fos}		{{FOS}}
\newcommand{\za}		{z_{\rm a}}
\newcommand{\ze}		{z_{\rm e}}
\newcommand{\kms}		{km~s$^{-1}$} 
\newcommand{\cmsq}		{cm$^{-2}$}
\newcommand{\NH}		{N({\rm H})}
\newcommand{\hst}       	{{\it HST\/}}
\newcommand{\phc}		{\phm{:}}

\lefthead{Papovich, Norman, Bowen, {\em et al.}}
\righthead{\hst\ Observations of The Associated Absorption Line Systems in \qso}

\begin{document}
\title{{\it Hubble Space Telescope} Observations of the Associated Absorption Line Systems in Q0122+0338\altaffilmark{1}}
\author{Casey Papovich\altaffilmark{2}, Colin A. Norman\altaffilmark{2,3}, David V. Bowen\altaffilmark{2,4}, Tim Heckman\altaffilmark{2}, \\ Sandra Savaglio\altaffilmark{3,5}, Anton M. Koekemoer\altaffilmark{3}, and J. Chris Blades\altaffilmark{3}}

\altaffiltext{1}{Based on observations with the NASA/ESA Hubble Space
Telescope, obtained at the Space Telescope Science Institute, which is
operated by the Association of Universities for Research in Astronomy,
Inc., under NASA contract No. NAS5-26555.}  \altaffiltext{2}{The Henry
A. Rowland Department of Physics and Astronomy, The Johns Hopkins
University, 3400 N. Charles Street, Baltimore, MD 21218}
\altaffiltext{3}{Space Telescope Science Institute, 3700 San Martin
Drive, Baltimore, MD 21218} \altaffiltext{4}{Current address:
Princeton University Observatory, Peyton Hall, Princeton, NJ
08544-1001} \altaffiltext{5}{On assignment from the Space Science
Department of the European Space Agency}

\begin{abstract}

We have studied a spectrum of \qso\ ($\ze \sim 1.202$) obtained by the
Faint Object Spectrograph (FOS) on board the {\it Hubble Space
Telescope} (\hst).  We present the analysis of three associated ($\za
\sim \ze$) absorption systems  at $z\:=\:1.207$, 1.199 and 1.166. The
most complex of these at $\za \sim 1.207$ shows strong absorption from
the highly ionized transitions of \lya, \lyb, \ion{N}{5}, \ion{O}{6},
\ion{Si}{3}, \ion{Si}{4}, and possibly \ion{P}{5}.  We derive
(minimal) ionic column densities for this system of $N($\ion{H}{1}$) =
10^{15.3}$~\cmsq, $N($\ion{N}{5}$) = 10^{14.8}$~\cmsq,
$N($\ion{O}{6}$) = 10^{15.4}$~\cmsq, $N($\ion{Si}{3}$) =
10^{13.3}$~\cmsq, and $N($\ion{Si}{4}$) = 10^{13.7}$~\cmsq.  By
comparing the derived column densities with those predicted from
numerical photoionization models, we find that conditions in the
absorbing gas are consistent with an absorber with a metallicity $\sim
2Z_\odot$ and a total absorbing column density of $\NH \simeq 2 \times
10^{19}$~\cmsq. The kinematics of the absorption lines in the $\za
\sim 1.207$ system suggest that a correlation exists between the
relative velocity and the creation ionization potential energy for
each transition.  This is evidence that a complex, multi-component
absorber exists.  Althought the location of the absorber is uncertain
(intrinsic versus intervening), we consider the origin of this
absorption system using the available data and discuss how the high-ionization, and
high-metallicity indicate that the absorber may be intrinsic to
\qso.

\end{abstract}

\section{Introduction}

Absorption lines in QSO spectra offer a direct means of tracing the
nature and distribution of gaseous matter in the Universe.  The
presence of metal lines in these systems ties the absorbing gas to the
processes of galaxy formation and evolution.
Associated absorption systems --- those with the absorption redshift,
$\za$, close to the QSO emission redshift, $\ze$ --- have been
classified separately from the general class of ($\za << \ze$)
narrow-line absorption systems, and their origin remains
uncertain.  These systems are characterized by highly ionized
absorption lines and are generally defined by the convention that the
relative velocity between the emission redshift and the absorption
redshift is $< 5000$~\kms\ (\cite{Weymann79}; \cite{Foltz86};
\cite{Anderson87}; \cite{Foltz88}).  Identifying the mechanism for
associated absorption has potentially important implications for
galactic formation and evolution.  If the associated systems originate
in regions local to the QSO, then they provide a direct probe into the
inner regions of active galaxies.

Associated systems may arise in the halos of neighboring galaxies or
the host galaxy of the QSO itself with a scale length of tens of
kiloparsecs.  Foltz et al.\ (1986) observed a correlation between
associated systems and radio-loud QSOs.  The fact
that the environments surrounding radio-loud QSOs tend to have more
nearby galaxies (\cite{Smith90}) may be construed as evidence
that $\za \sim \ze$ systems originate in neighboring galactic
structures.  However, Anderson et al.\ (1987) report that although a
high incidence of associated systems occur in radio-loud QSOs, there
is no statistically significant correlation between the two.  Barthel,
Tytler, \& Thomson (1990) combined a study with previous work
(\cite{Barthel88}) and found $\za \sim \ze$ systems biased towards
neither radio-loud nor radio-quiet QSOs.

Some associated absorption systems show evidence of being
intrinsic to the QSO engine.  These systems are typically highly
ionized given the presence of lithium--like \ion{O}{6}, \ion{N}{5},
and \ion{C}{4} which implies the gas is photoionized by the QSO.  In
several studies with high quality data, the metallicities of these
systems have been solar to several times solar (\cite{Wampler93};
\cite{Moller94}; \cite{Petitjean94}; \cite{Hamann97};
\cite{Hamann97b}).  If these systems originate near the central
regions of the QSO host galaxy,
then the high metallicities agree with predictions of galactic
chemical evolution (\cite{Hamann93}; \cite{Matteucci93};
\cite{Hamann97d}; see Hamann \& Ferland 1999 for a recent review) and with
the metallicities of QSO environments derived from studies of broad 
absorption line (BAL) systems and QSO emission lines (\cite{Hamann93}; 
\cite{Ferland96b}; \cite{Turnshek96}; \cite{Hamann97}).  However, the 
poorly understood effects of the covering fraction of the source by 
the absorber, and inherent line saturation effects produce systematic 
uncertainties in converting the observed column densities to 
metallicities. 

More evidence for a class of intrinsic, narrow-lined absorption
systems comes from dynamical observations.  Studies of some BAL and
associated systems show time variability in their absorption profiles.
Assuming the QSO is the ionizing source, time variability in the
absorber implies either that a variable QSO flux has changed the
ionization state of the cloud, or that the density has changed due to
motions not along the line of sight.  Either scenario requires small
distance scales (\cite{Barlow92}).  Multi-epoch observations of
Q0835+5804 and Q1157+0128 by Aldcroft, Bechtold, \& Foltz (1997) show
line strength variability in intrinsic absorption features on the time
scale of a few years. In Q2343+125, Hamann et al.\ (1997a) observe
temporal variations in the $\za \sim \ze$ absorber between two epochs
separated by $\sim 1$ yr.  Furthermore, they derive covering factors
less than unity which indicates that the absorber does not completely
occult the emission source.  Partial coverage of small, dense
absorbers produce shallow absorption features with multiplet line
equivalent width strength ratios of order unity (Petitjean et al.\
1994; \cite{Hamann95}; \cite{Barlow97}; \cite{Hamann97c}). Both
multi-epoch variations and partial covering factors give strong
support to the idea that the absorption originates near the QSO
emission source.

\qso\ is a radio-quiet QSO (\cite{Veron98}) at $\ze = 1.202$, with
magnitude, $V = 17.7$ ($M_V \approx -27$).  Arp \& Duhalde (1985)
first observed \qso\ as part of a search for quasars in the field of
NGC 520 (their object 40), and the QSO was used by Norman~et~al.\
(1996) to search for UV absorption lines from the outflow of the
galaxy.  X-ray surveys by Yuan et al.\ (1998) show \qso\ to be a
typical radio-quiet QSO with a steep-spectrum slope and an X-ray flux,
$F_X(0.1-2.4\;{\rm keV}) = 1.3\times10^{-13}$ erg~\cmsq~s$^{-1}$.

In this paper, we report the detection of three associated systems in
\qso\ at $\za \sim 1.166$, 1.199, and 1.207.  The corresponding
relative velocities to the emission redshift are $\simeq -5000$,
$-410$, and $+680$ \kms, respectively.  The relative velocity is
defined as
\begin{equation}
\frac{\Delta v}{c} = \frac{(1+\za)^2 - (1+\ze)^2 }{(1+\za)^2 +
(1+\ze)^2}.
\end{equation}
such that a negative relative velocity indicates motion towards us
with respect to the emission redshift.

In \S2, we discuss the procedure used to obtain and reduce the
\hst/\fos\ data.  In \S 3, we present the analysis of the absorption
lines including the line detection procedure and the 
column density derivations.  In \S 4, we describe the characteristics
of each of the three associated absorption systems.  Using arguments
based on ionization, metallicity, and kinematics, we constrain the
origin of each system.

\section{Data Acquisition and Reduction}

\begin{figure*}
\figurenum{1}
\epsscale{2.0}
\plotone{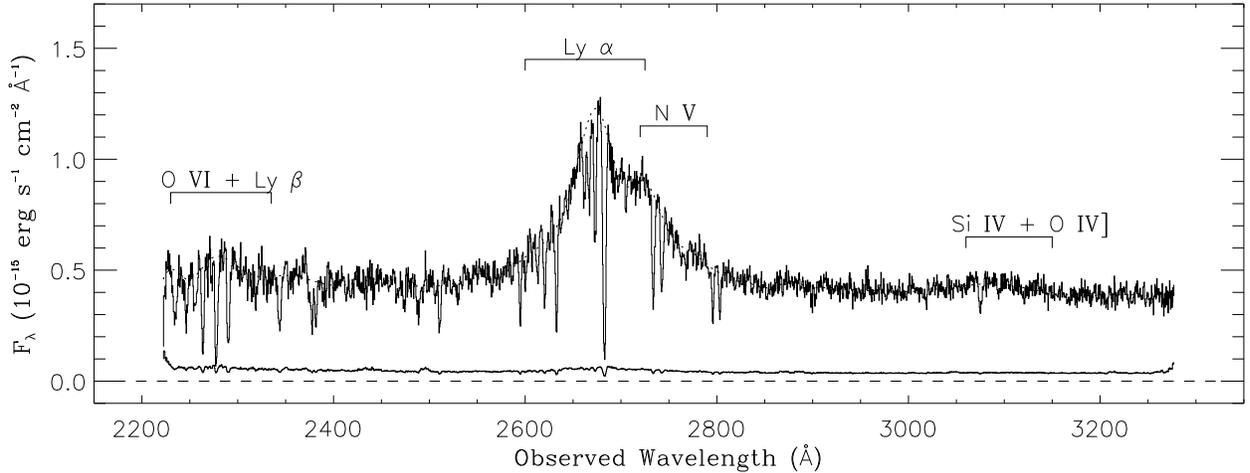}
\caption{The \hst/\fos\ spectrum of \qso\ from the G270H grating.  The spectrum has a resolution of FWHM $\sim 225$~\kms.  The solid lines show the flux and the $1\sigma$ errors and the dotted line shows the unabsorbed continuum.  Prominent emission features are labeled.\label{Spectrum}}
\end{figure*}

The spectrum of \qso\ was taken by Norman~et~al.\ (1996) with the Faint
Object Spectrograph (\fos) and pre-COSTAR optics on 9-Sept-1993 and
22-Nov-1993. The observations were obtained using the G270H grating
with the Red Digicon detector through the $0\farcs25\times 2\farcs0$ 
ACCUM mode.  The observed wavelength range
is $\Delta\lambda = 2222.3 - 3277.2$~\AA\ with a dispersion of
0.51~\AA\ pixel$^{-1}$ and a corresponding resolution of
$187-275$~\kms.  Each diode was sampled by four (NXSTEPS=4) pixels with
a total of five (OVERSCAN=5) overscan steps.  The total exposure time
was 5760 seconds (96 minutes).

\setcounter{footnote}{0}

The general method used to handle the data reduction is outlined by the QSO
Absorption Line Key Project (QALKP; e.g., \cite{Bahcall93}).  Reduction of the
data was performed using
the IRAF/STSDAS\footnote{ The Image Reduction and
Analysis Facility (IRAF) software is provided by the National Optical
Astronomy Observatories (NOAO), which is operated by the Association of
Universities for Research in Astronomy, Inc., under contract to the National
Science Foundation.  The Space Telescope Science Data Analysis System (STSDAS)
is distributed by the Space Telescope Science Institute} standard pipeline
software and the datasets were coadded using standard IRAF routines. The
photon noise errors and data quality were calculated and propagated through the
pipeline calibration.  We utilize a wavelength scale in which the wavelengths
of the absorption due to the Galactic \ion{Mg}{2} $\lambda\lambda 2796$, 2803
doublet equal the vacuum wavelengths of Verner, Bathel \& Tytler (1994).  A
wavelength shift of $\Delta\lambda = +0.425$~\AA\ was implemented to remove
the offset between them.

We fit the spectrum continuum (including the broad emission lines) using the
STSDAS package {\tt FIT1D} following methods outlined by the QALKP
(\cite{Schneider93}).  The spectrum was fit automatically using a $\chi^2$
minimization of a series of cubic spline curves with interactive manipulation
in spectral regions that are poorly fit. The fit was deemed satisfactory after we
ensured that residuals in the normalized spectrum were comparable to the
Poisson noise.  In Figure~\ref{Spectrum}, we present the flux and
$1\sigma$ errors versus the observed wavelength. Normalized flux and error
files were generated by dividing the flux and error files from the pipeline by
the fitted continuum.

\section{Detection and Analysis of Absorption Lines}

\subsection{Absorption Line Identification}

We employed an automatic, objective search for statistically
significant absorption features in the normalized spectrum using the
algorithm developed for the QALKP (\cite{Schneider93}). The
line-finding algorithm searches for minimum feature widths based on
the \fos\ line spread function for this grating and aperture ($\sim
1.95$~\AA).  We restricted our line search to features with
statistical {\em significance level} [$SL \equiv W_{\rm
obs}/\sigma(W)$] $SL \geq 2.6$ which corresponds to a detection of
weak lines with observed equivalent width $W_{\rm obs} \gtrsim
0.25$~\AA.

\placetable{assoc_table}
\begin{deluxetable}{lrcccrrrcrr}
\tablewidth{0pt} \tablefontsize{\scriptsize} \tablewidth{0pt} \tablecolumns{11}
\tablecaption{Absorption lines identified in
Q0122+0338\label{assoc_table}} \tablehead{\colhead{} &
\colhead{$\lambda_{\rm obs}$} & \colhead{$\sigma(\lambda)$} &
\colhead{$W_{\rm obs}$} & \colhead{$\sigma(W)$} & \colhead{} &
\colhead{$\lambda_{\rm fit}$} & \colhead{$W_{\rm fit}$} &
\colhead{${\rm FWHM}_{\rm fit}$} & \colhead{} & \colhead{} \\
\colhead{} & \colhead{(\AA)} & \colhead{(\AA)} & \colhead{(\AA)} &
\colhead{(\AA)} & \colhead{SL} & \colhead{(\AA)} & \colhead{(\AA)} &
\colhead{(km s$^{-1}$)} & \colhead{ID} & \colhead{$\za$} } \startdata
1 & 2234.94 & 0.17 & 1.40 & 0.14 & 9.7 & $ 2234.68 \pm 0.16 $ & $ 1.97\pm 0.22 $ & $ 517\pm  62\phn $ & O VI 1031.93\phc & \phs 1.16554 \\
2 & 2246.78 & 0.22 & 1.16 & 0.15 & 7.6 & $ 2247.01 \pm 0.16 $ & $ 1.54\pm 0.19 $ & $ 517\pm  62\phn $ & O VI 1037.62\phc & \phs 1.16554 \\
3 & 2255.91 & 0.27 & 0.90 & 0.15 & 5.9 & $ 2255.98 \pm 0.29 $ & $ 1.13\pm 0.18 $ & $ 444\pm  68\phn $ & H I 1025.72\phc & \phs 1.19941 \\
4 & 2264.02 & 0.11 & 2.08 & 0.11 & 18.6 & $ 2264.03 \pm 0.08 $ & $2.23 \pm 0.14 $ & $ 355 \pm  26\phn $ & H I 1025.72\phc & \phs 1.20725\\  
5 & 2273.67 & 0.49 & 0.46 & 0.15 & 3.1 & $ 2273.33 \pm 0.16 $ & $0.29 \pm 0.13 $ & $ 108 \pm  30\phn $ & \nodata & \nodata \\  
6 &2278.05 & 0.09 & 2.56 & 0.10 & 25.7 & $ 2278.03 \pm 0.06 $ & $ 3.07\pm 0.14 $ & $ 418 \pm  19\phn $ & O VI 1031.93\phc & \phs 1.20755 \\
7 & 2290.59 & 0.10 & 2.07 & 0.10 & 20.5 & $ 2290.59 \pm 0.06 $ & $2.48 \pm 0.12 $ & $ 418 \pm  19\phn $ & O VI 1037.62\phc & \phs1.20755 \\  
8 & 2315.15 & 0.52 & 0.47 & 0.16 & 2.9 & $ 2315.06 \pm0.18 $ & $ 0.34 \pm 0.13 $ & $ 140 \pm  65\phn $ & \nodata & \nodata\\  
9 & 2319.22 & 0.32 & 0.66 & 0.14 & 4.7 & $ 2319.20 \pm 0.27 $ & $0.70 \pm 0.14 $ & $ 312 \pm  64\phn $ & \nodata & \nodata \\  
10 &2331.08 & 0.54 & 0.41 & 0.15 & 2.8 & $ 2330.87 \pm 0.19 $ & $ 0.31 \pm0.12 $ & $ 145 \pm  70\phn $ & \nodata & \nodata \\  
11 & 2344.12 &0.14 & 1.51 & 0.12 & 13.0 & $ 2344.05 \pm 0.15 $ & $ 2.00 \pm 0.16 $ &$ 470 \pm  42\phn $ & Fe II 2344.21\phc & $-0.00007$ \\  
12 & 2377.95& 0.17 & 1.25 & 0.12 & 10.3 & $ 2377.87 \pm 0.18 $ & $ 1.47 \pm 0.26 $& $ 378 \pm  93\phn $ & Fe II 2374.46: & \phs 0.00144 \\  
13 & 2381.88& 0.18 & 1.17 & 0.12 & 9.5 & $ 2382.00 \pm 0.14 $ & $ 1.09 \pm 0.15 $& $ 269 \pm  42\phn $ & Fe II 2382.76\phc & $-0.00032$ \\  
14 &2389.86 & 0.55 & 0.38 & 0.14 & 2.7 & $ 2390.26 \pm 0.65 $ & $ 0.55 \pm0.17 $ & $ 492 \pm 109 $ & \nodata & \nodata \\  
15 & 2466.86 & 0.53 &0.36 & 0.13 & 2.8 & $ 2466.92 \pm 0.55 $ & $ 0.48 \pm 0.14 $ & $ 468\pm  42\phn $ & P V 1117.98: & \phs 1.20659 \\  
16 & 2474.00 & 0.32 &0.61 & 0.12 & 4.8 & $ 2474.03 \pm 0.26 $ & $ 0.52 \pm 0.11 $ & $ 194\pm  32\phn $ & \nodata & \nodata \\  
17 & 2488.98 & 0.23 & 0.87 &0.13 & 6.9 & $ 2488.96 \pm 0.26 $ & $ 1.12 \pm 0.21 $ & $ 406 \pm 103$ & \nodata & \nodata \\  
18 & 2506.65 & 0.57 & 0.33 & 0.13 & 2.6 & $2506.45 \pm 0.37 $ & $ 0.27 \pm 0.14 $ & $ 189 \pm 127 $ & \nodata &\nodata \\  
19 & 2511.17 & 0.15 & 1.22 & 0.11 & 11.5 & $ 2511.13 \pm0.11 $ & $ 1.23 \pm 0.11 $ & $ 274 \pm  25\phn $ & \nodata & \nodata\\  
20 & 2530.00 & 0.35 & 0.51 & 0.12 & 4.4 & $ 2529.97 \pm 0.33 $ & $0.58 \pm 0.12 $ & $ 331 \pm  63\phn $ & \nodata & \nodata \\  
21 &2565.28 & 0.48 & 0.35 & 0.11 & 3.1 & $ 2565.46 \pm 0.26 $ & $ 0.32 \pm0.10 $ & $ 192 \pm  74\phn $ & \nodata & \nodata \\  
22 & 2595.22 &0.13 & 1.24 & 0.09 & 14.2 & $ 2595.26 \pm 0.07 $ & $ 1.14 \pm 0.09 $ &$ 220 \pm  19\phn $ & \nodata & \nodata \\  
23 & 2600.59 & 0.25 & 0.59& 0.10 & 6.1 & $ 2600.61 \pm 0.18 $ & $ 0.56 \pm 0.11 $ & $ 234 \pm67\phn $ & Fe II 2600.17\phc & \phs 0.00017 \\  
24 & 2613.37 & 0.24 &0.58 & 0.09 & 6.3 & $ 2613.45 \pm 0.18 $ & $ 0.54 \pm 0.08 $ & $ 231\pm  35\phn $ & Si III 1206.50\phc & \phs 1.16614 \\  
25 & 2620.84 &0.12 & 1.12 & 0.08 & 13.9 & $ 2620.83 \pm 0.09 $ & $ 1.11 \pm 0.09 $ &$ 257 \pm  25\phn $ & \nodata & \nodata \\  
26 & 2633.10 & 0.08 & 1.76& 0.06 & 27.5 & $ 2633.13 \pm 0.05 $ & $ 1.78 \pm 0.08 $ & $ 270 \pm14\phn $ & H I 1215.67\phc & \phs 1.16599 \\  
27 & 2662.38 & 0.13 &0.80 & 0.06 & 12.5 & $ 2662.33 \pm 0.14 $ & $ 0.91 \pm 0.10 $ & $ 337\pm  47\phn $ & Si III 1206.50\phc & \phs 1.20665 \\  
28 & 2667.00 &0.11 & 0.91 & 0.06 & 15.4 & $ 2667.07 \pm 0.12 $ & $ 1.03 \pm 0.10 $ &$ 330 \pm  40\phn $ & \nodata & \nodata \\  
29 & 2673.39 & 0.07 & 1.39& 0.05 & 26.6 & $ 2673.38 \pm 0.06 $ & $ 1.56 \pm 0.07 $ & $ 317 \pm16\phn $ & H I 1215.67\phc & \phs 1.19910 \\  
30 & 2683.14 & 0.04 &2.80 & 0.03 & 91.6 & $ 2683.15 \pm 0.03 $ & $ 3.71 \pm 0.07 $ & $ 415\pm   8\phn\phn $ & H I 1215.67\phc & \phs 1.20714 \\  
31 & 2693.63 &0.35 & 0.34 & 0.08 & 4.3 & $ 2693.38 \pm 0.56 $ & $ 0.58 \pm 0.12 $ &$ 583 \pm 103 $ & \nodata & \nodata \\  
32 & 2705.67 & 0.51 & 0.23 &0.08 & 3.0 & $ 2705.99 \pm 0.10 $ & $ 0.13 \pm 0.05 $ & $ 189 \pm58\phn $ & \nodata & \nodata \\  
33 & 2734.14 & 0.07 & 1.63 & 0.06 &27.3 & $ 2734.17 \pm 0.05 $ & $ 1.88 \pm 0.07 $ & $ 327 \pm  13\phn $& N V 1238.82\phc & \phs 1.20708 \\
34 & 2743.04 & 0.10 &1.31 & 0.07 & 19.0 & $ 2742.96 \pm 0.05 $ & $ 1.50 \pm 0.07 $ & $ 327\pm  13\phn $ & N V 1242.80\phc & \phs 1.20708 \\  
35 & 2763.37 & 0.44& 0.32 & 0.10 & 3.3 & $ 2763.32 \pm 0.46 $ & $ 0.34 \pm 0.11 $ & $ 285\pm  96\phn $ & \nodata & \nodata \\  
36 & 2768.35 & 0.30 & 0.45 &0.09 & 4.9 & $ 2768.38 \pm 0.35 $ & $ 0.55 \pm 0.17 $ & $ 350 \pm 171$ & \nodata & \nodata \\  
37 & 2796.26 & 0.14 & 1.14 & 0.09 & 13.0 & $2796.27 \pm 0.09 $ & $ 1.09 \pm 0.09 $ & $ 224 \pm  23\phn $ & Mg II 2796.35\phc & $-0.00003$ \\  
38 & 2803.62 & 0.16 & 0.98 & 0.09 & 10.5& $ 2803.59 \pm 0.10 $ & $ 0.92 \pm 0.10 $ & $ 216 \pm  27\phn $ & Mg II 2803.53\phc & \phs 0.00002 \\  
39 & 2819.45 & 0.56 & 0.28 & 0.11 &2.6 & $ 2819.48 \pm 0.51 $ & $ 0.28 \pm 0.13 $ & $ 240 \pm 162 $ &\nodata & \nodata \\  
40 & 2899.70 & 0.41 & 0.41 & 0.11 & 3.7 & $2899.81 \pm 0.26 $ & $ 0.39 \pm 0.12 $ & $ 188 \pm  72\phn $ & \nodata& \nodata \\  
41 & 2903.07 & 0.53 & 0.32 & 0.11 & 2.8 & $ 2903.11 \pm0.42 $ & $ 0.23 \pm 0.10 $ & $ 244 \pm  88\phn $ & \nodata & \nodata\\  
42 & 3018.67 & 0.53 & 0.32 & 0.12 & 2.8 & $ 3018.95 \pm 0.16 $ & $0.32 \pm 0.08 $ & $ 212 \pm 132 $ & Si IV 1393.76\phc & \phs 1.16605\\  
43 & 3075.42 & 0.20 & 0.80 & 0.10 & 7.8 & $ 3075.46 \pm 0.20 $ & $0.93 \pm 0.11 $ & $ 291 \pm  34\phn $ & Si IV 1393.76\phc & \phs1.20660 \\  
44 & 3094.61 & 0.56 & 0.29 & 0.11 & 2.6 & $ 3095.35 \pm0.20 $ & $ 0.29 \pm 0.11 $ & $ 291 \pm  34\phn $ & Si IV 1402.77\phc &\phs 1.20660 \\  
45 & 3120.90 & 0.46 & 0.38 & 0.12 & 3.3 & $ 3120.82\pm 0.60 $ & $ 0.52 \pm 0.14 $ & $ 379 \pm  78\phn $ & \nodata &
\nodata \\

\enddata
\end{deluxetable}

\begin{figure*}
\figurenum{2}
\epsscale{1.0}
\plotfiddle{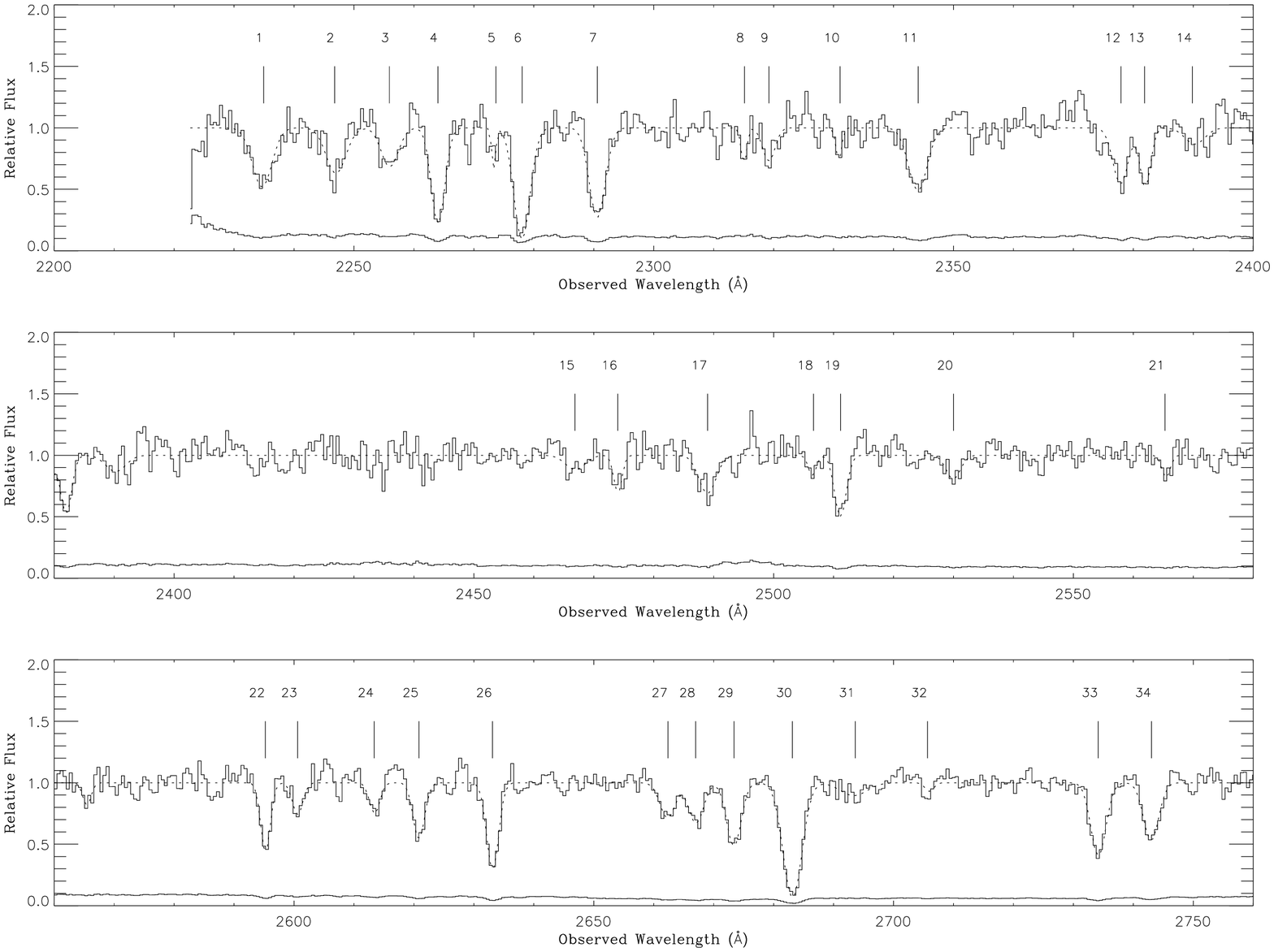}{7in}{90}{75}{75}{216}{0}
\caption{The normalized spectrum of \qso.  The solid line shows the normalized spectrum, and the dotted line is the best fit from {\tt SPECFIT} to the absorption features detected by the line search algorithm.  Absorption lines are labeled with numbers corresponding to the first column of Table~\ref{assoc_table}.  The normalized, $1\sigma$ error from the data is drawn as a solid line below the spectrum.  \label{NormalizedSpectrum}}
\end{figure*}

All absorption lines detected were fit with the IRAF package {\tt
SPECFIT} (\cite{Kriss94}).  The profiles of the absorption lines are
modeled with single component Gaussian profiles which is appropriate
for marginally resolved absorption features since the centroid of the
\fos\ line spread function is best fit by a Gaussian (Evans 1993).
{\tt SPECFIT} best fits all the specified features by minimizing
$\chi^2$ with an optimal Marquardt routine.  Where applicable, we
improved the fit by requiring the identified members of line doublets
to have identical redshifts and FWHM.  We list in columns 2-6 of Table~\ref{assoc_table} the wavelength, equivalent width (along with
$1\sigma$ errors) and the significance level for each line detected by
the search algorithm.  The wavelength, equivalent width and FWHM
derived by {\tt SPECFIT} are presented in columns 7-9 designated by
the subscript FIT.  Line identifications and redshifts are presented
in columns 10 and 11.  Figure~\ref{NormalizedSpectrum} shows the
normalized spectrum with the normalized continuum and fitted
absorption lines overdrawn. The normalized, $1\sigma$ errors to the
data are also shown.  We list in the first column of Table~\ref{assoc_table} the number assigned to each absorption line by the
detection algorithm.  These numbers are used to label the
corresponding absorption line in Figure~\ref{NormalizedSpectrum}.

\placefigure{NormalizedSpectrum}

\begin{figure*}
\figurenum{2 cont}
\epsscale{1}
\plotfiddle{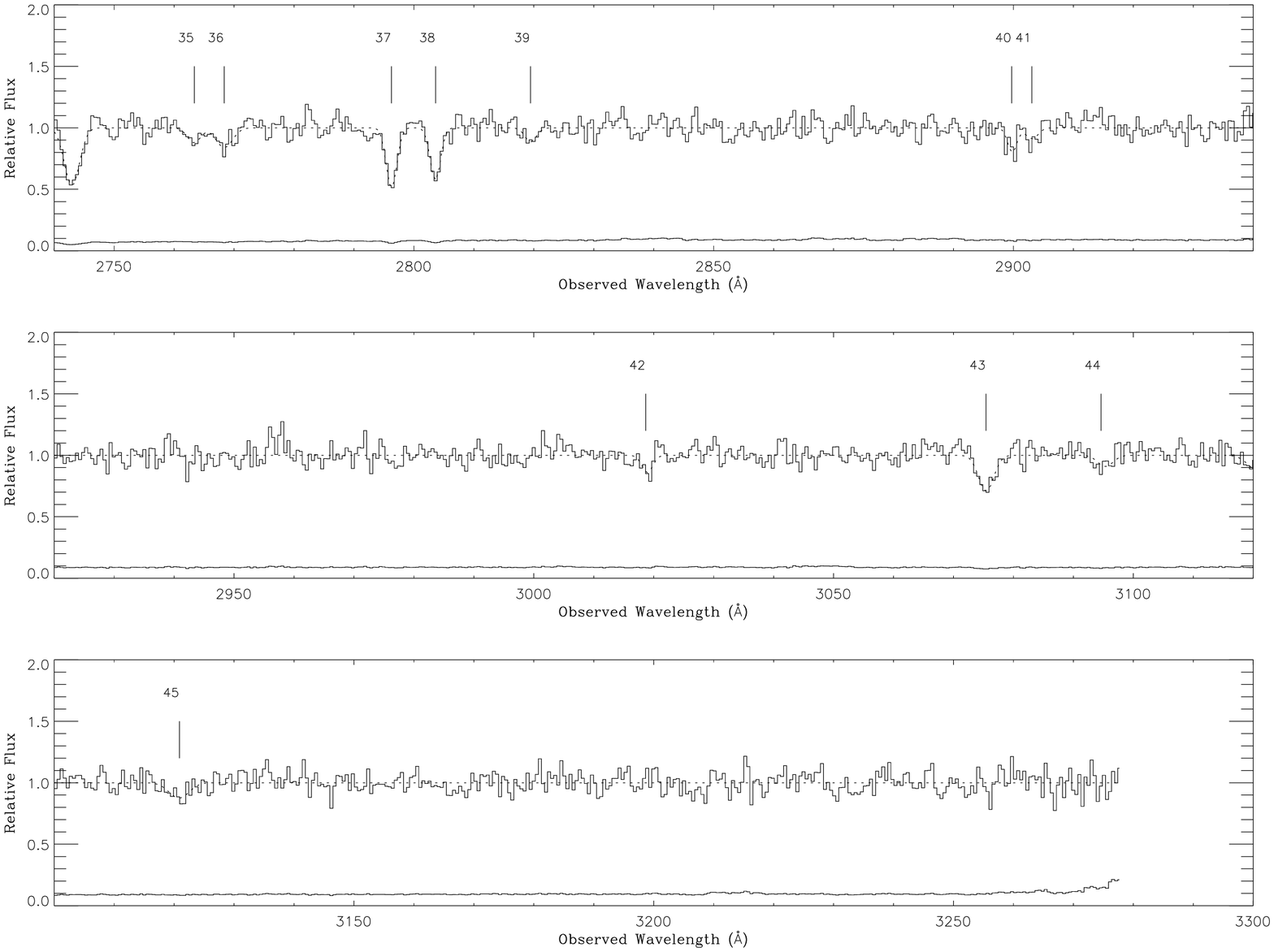}{7in}{90}{75}{75}{216}{0}
\caption{}
\end{figure*}

Our detection threshold $SL \geq 2.6$ is below the ``complete sample''
threshold used by the QALKP, $SL > 4.5$.  Since a major goal of the
Key Project was the statistical distribution of \lya\ absorbers, they
considered only spectral features above this threshold to ensure no
spurious detections.  However, in the Bahcall et al.\ (1993)
discussion of individual systems, they include lines of significance
level $3.0 \leq SL \leq 4.5$ when such lines are expected (their
``incomplete sample'').  Our choice includes the detection of the
weaker doublet member \ion{Si}{4} $\lambda 1402$ of the $\za = 1.207$
system which we expect given that the stronger doublet member,
\ion{Si}{4} $\lambda 1393$, is present with $SL \simeq 8$.  Therefore,
we include lines with $SL \geq 2.6$ in Table~\ref{assoc_table} for
illustrative reasons, but we use lines with $SL \geq 3.0$ when
focusing on the properties of individual systems.

\subsection{Optical Depths and Column Densities of Absorption Lines}

Analyses of the absorption lines suffer from inherent problems arising
from uncertainties in the pre-COSTAR \fos\ Line Spread Function (LSF).
Our \hst/\fos\ spectra have only moderate resolution ($\sim 225$~\kms)
and the convolution of the line widths with the LSF broadens the FWHM
of the line which can hide multiple line components.  Both of these
limitations tend to underestimate the optical depth (and hence column
density) of absorption lines in the standard {\em curve-of-growth}
analysis.

\setcounter{figure}{2}
\begin{figure*}
\figurenum{3}
\epsscale{0.5}
\plotfiddle{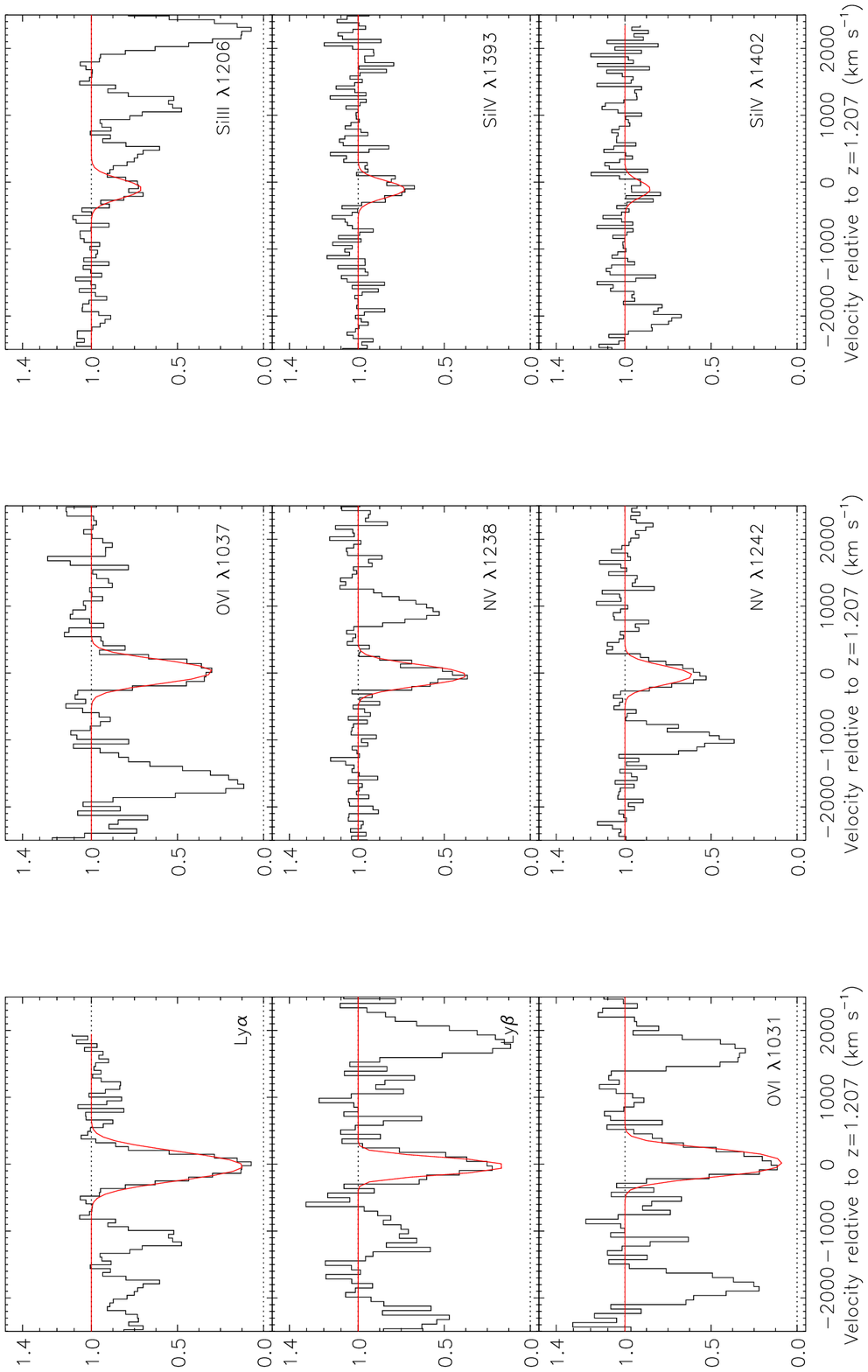}{3.95in}{270}{75}{75}{-300}{366}
\caption{The line profiles for the nine absorption lines of the $\za \sim 1.207$ system are shown as normalized flux versus the velocity (\kms) relative to $\za = 1.207$.  The best fit from {\tt MADRIGAL} absorption line profiles are overdrawn as a gray line.  The line identifications are shown in each panel. \label{line_profiles}}
\end{figure*}

Therefore, we derived velocities relative to $z_a$, $\Delta v$,
Doppler parameters, $b$, and column densities, $N(X_i)$, using the
line profile fitting routine {\tt MADRIGAL} which compares observed
absorption line profiles with theoretical models (\cite{Bowen95}).
Unlike {\tt SPECFIT}, {\tt MADRIGAL} uses the oscillator strength for
a given transition and its use is only appropriate for the identified
absorption lines.  Initial line profiles were computed given initial
values for $b$ and $N(X_i)$ using the equivalent widths and FWHM of
the {\tt SPECFIT} results. These were convolved with the LSF of the
pre-{\tt COSTAR} $0\farcs25\times 2\farcs0$ aperture, which is  a
Gaussian profile with a FWHM of 3.36 pixels\footnote{ see {\tt
http://www.stsci.edu/ftp/instrument\_news/FOS/fos\_curstat.html}}.  A
final set of values for $\Delta v$, $b$ and $N(X_i)$ were derived by
varying the initial parameters and minimizing a $\chi^2$ statistic
between the model and the observed data.  In all cases the data is of
insufficient resolution to differentiate between single and
multi--component absorption line profiles and the best fits are
obtained with single component absorption lines.  As a result, the
derived values for the Doppler parameter indicate the spread of
observed line profile and it is more useful to report the FWHM of the
lines.\footnote{For purely Gaussian profiles, ${\rm FWHM} = \sqrt{4
\ln 2}\; b$.}  For the three $\za \sim \ze$ systems, the best values
for $\Delta v$, $N(X_i)$ and FWHM are shown in Table~\ref{column_depth_table}.

\begin{deluxetable}{lrcccccc}
\tablewidth{0pt} \tablefontsize{\small}
\tablecolumns{8}
\tablecaption{Derived Column Densities and Doppler Parameters for 
Identified Absorption Lines in Q0122+0338\label{column_depth_table}}
\tablehead{\colhead{} & \colhead{Creation} & \colhead{} & \colhead{} &
\colhead{} & \colhead{} & \colhead{} & \colhead{} \\
\colhead{} & \colhead{Energy} & \colhead{$\Delta v\tablenotemark{a}$} &
\colhead{$\sigma(\Delta v)$\tablenotemark{b}} & \colhead{FWHM} & 
\colhead{$\sigma({\rm FWHM})$\tablenotemark{b}} & \colhead{$\log N(X_i)$\tablenotemark{c}} & 
\colhead{$\sigma( \log N )$\tablenotemark{b}} 
\\ 
\colhead{Line} & \colhead{(eV)} & \colhead{(km s$^{-1}$)} & 
\colhead{(km s$^{-1}$)} & \colhead{(km s$^{-1}$)} & 
\colhead{(km s$^{-1}$)} & \colhead{(cm$^{-2}$)} & \colhead{(cm$^{-2}$)}
}
\startdata
\cutinhead{$\za = 1.166$}
\ion{O}{6}  & 113.900 & $-60$       & 20     & 435     & \phn 57    & 15.0 & 0.04 \\
\ion{Si}{3} & 16.346  & \phs 20     & 28     & 125     & \phn 93    & 13.2 & 0.28 \\
\ion{H}{1}  & 0.000   & \phn $-2$   & \phn 5 & 157     & \phn 93    & 14.5 & 0.01 \\
\ion{Si}{4} & 33.494  & \phn \phs 9 & 35     & \phn 48 & \phn 85    & 13.3 & 0.02 \\

\cutinhead{$\za = 1.199$}
\ion{H}{1} & 0.000    & \phs 18     & 19     & \phn 83 & 120        & 14.6 & 0.04 \\

\cutinhead{$\za = 1.207$}
\ion{O}{6}  & 113.900 & \phs 77     & \phn 7 & 325     & \phn 15    & 15.4 & 0.02 \\
\ion{Si}{3} & 16.346  & $-50$       & 13     & 143     & \phn 28    & 13.3 & 0.04 \\
\ion{H}{1}  & 0.000   & \phs 24     & \phn 3 & 215     & \phn\phn 4 & 15.3 & 0.03 \\
\ion{N}{5}  & 77.474  & \phs 13     & \phn 6 & 295     & \phn 13    & 14.8 & 0.02 \\
\ion{Si}{4} & 33.494  & $-65$       & 54     & 273     & \phn 72    & 13.7 & 0.12
\enddata
\tablenotetext{a}{Here, the relative velocity is computed with respect to the absorption system redshift for each system.}
\tablenotetext{b}{These uncertainties represent purely statistical standard deviations and do not include systematic errors due to effects such as proper normalization or extraneous line contamination.} 
\tablenotetext{c}{Due to the low resolution of the FOS, the derived column densities are lower limits.  See text for discussion.}
\end{deluxetable}

We estimated the uncertainties on the derived values of $\Delta v$,
$N(X_i)$ and $b$ by using a Monte--Carlo analysis of the spectrum.
Using the same procedure as above, we measured the column densities
and Doppler parameters in each of $\simeq 100$ artificial spectrum
which were constructed with Poisson noise identical to that of the FOS
data.  The new values differ from the original parameters by $\Delta
v$, $\Delta b$, and $\Delta N$ and the distribution of these
parameters is approximately Gaussian in one dimension.  The $\sigma(
N)$ and $\sigma(b)$ are correlated and the strength of this is
dependent on the width of the line.  For example, narrow lines which
cover only a few pixels are more sensitive to changes as a result of
Poisson noise and therefore have greater uncertainties for $N(X_i)$.
We adopt the $\sigma$ of the one dimensional $\Delta N(X_i)$--$\Delta
b$ distribution as the $1\sigma$ uncertainty.  In general, however,
the $+\sigma$ uncertainty of one parameter should be taken with
$-\sigma$ in the other due to the correlation.  These errors only
represent the Poisson statistics and do not incorporate systematic
effects due to the unknown multi--component composition of the
absorption lines and uncertainties in the calibration.

Any unresolved, narrow absorption lines contribute little to the
equivalent width but may contain large column densities.  Therefore,
the derived values of $N(X_i)$ are, strictly speaking, lower limits.
Since most of the systems appear to be highly ionized, the absorption
lines may be intrinsically broad.  In that case the measured column
densities are likely to be fairly close to the true values.
High--resolution spectra are required to differentiate any unresolved
components.

\subsection{Absorption Lines in the \lya\ Forest}

The analysis of the absorption lines in the range $2650-2700$~\AA\
requires special consideration since this region is strongly affected
by the \lya\ forest and is near the peak of the \lya\ emission line.
We have normalized the emission feature using the observed flux and
errors, but the true peak in the emission flux may be significantly
higher and strongly attenuated by the absorption lines.   Equivalent
widths derived for absorption lines in the region of the peak
(including: \ion{Si}{3}, $\za \sim 1.207$; \lya, $\za \sim 1.199$;
\lya, $\za \sim 1.207$) are underestimated if this is the case.
However, since the theoretical fit reproduces \lya\ and \lyb\ of the
$\za \sim 1.207$ system very well (cf.\ Figure~\ref{line_profiles}),
the attenuation cannot be too severe.

Intervening \lya\ lines may conspire to produce spurious detections or
increase the equivalent width of a detected line.  We use the
statistical distribution of \lya\ lines to predict the extent to which
this effect interferes with the analysis.  Weymann et al.\ (1998)
measure the differential number of \lya\ forest lines as a function of
redshift and rest equivalent width for QSO lines of sight.  Using the
empirical results of their sample, we expect $\sim 8$ \lya\ forest
lines with $SL > 4.5$ in our spectral range ($z_{\rm Ly\alpha} \simeq
0.83-1.18$), and we detect seven such lines (see Table~\ref{assoc_table}).  Therefore, these lines are consistent with
absorption due to the \lya\ forest.

\subsection{Relative Velocities and Redshifts}

The emission redshift of \qso\ was derived by Arp \& Duhalde (1985)
from observations of the \ion{C}{3}] $\lambda 1909$ and \ion{Mg}{2}
$\lambda 2800$ emission lines.  The reported emission redshift, $\ze =
1.202$, is the average from observations of separate nights, $\ze =
1.199$ and 1.205.  However, the relation between the redshift derived
from emission lines and the true QSO systemic redshift remains uncertain.
Tytler \& Fan (1992) demonstrated the statistical existence of
velocity shifts in QSO emission lines from the true systemic velocity.
The bias velocity shifts for the \ion{C}{3}] $\lambda 1909$ and
\ion{Mg}{2} $\lambda 2800$ emission lines (the lines used to determine
the redshift of \qso) are statistically blueshifted by $\sim
100$~\kms.  Similarly, Espey (1993) cites evidence that although the
high-ionization emission lines (e.g., \lya, \ion{C}{4} $\lambda 1550$)
are typically blueshifted with respect to forbidden lines (e.g.,
[\ion{O}{3}] $\lambda 5007$) by $\sim 1000$~\kms, the velocity
difference between the forbidden lines and the Balmer series and
low-ionization lines (e.g., \ion{Mg}{2} $\lambda 2800$) are
approximately nil.  Therefore, a fair estimate for the systemic
redshift of \qso\ is probably $\ze = 1.202\pm0.003$.

The systemic redshift has important implications for the relation
between the $\za \sim 1.207$ absorption system and the QSO host
galaxy.  The velocity shift between them ranges from slow to
moderately rapid infall ($270 \lesssim \Delta v \lesssim 1100$~\kms)
depending on the systemic redshift adopted.  If the velocity
difference is small, then the absorbing gas may be essentially at rest
with respect to the QSO and may be produced in the halo of the host
galaxy or a close neighboring galaxy.  If the velocity difference is
large, then either a substantial flow exits, or there are large peculiar
velocity dispersions among neighbors, or the absorber is moving as a
result of the cosmological expansion.

\subsection{Galactic Absorption Features}

We observe Galactic absorption in the spectrum of \qso\ from
intervening \ion{Fe}{2} $\lambda 2344,$ $\lambda 2382$, $\lambda
2600$, and \ion{Mg}{2} $\lambda\lambda 2796,$ 2803.\footnote{ We
observe a strong absorption feature  with $\lambda = 2595$~\AA\ and
$W_{\rm obs} \sim 1$~\AA\ which corresponds to Galactic Mn~II $\lambda
2594$.  Based on the oscillator strengths, this ion is expected to
produce the a comparable transition, Mn~II $\lambda 2576$.  Since the
latter is absent in the FOS spectrum, this absorption line is most
plausibly due to a chance intervening line (probably \lya).}   The
observed features are listed in Table~\ref{assoc_table}.  The lines of
the \ion{Mg}{2} $\lambda\lambda 2796,$ 2803 doublet have equivalent
widths $\sim 1$~\AA\ and an equivalent width ratio of $\sim$ 1:1.
Therefore, these lines are probably saturated and are comparable to
results from previous studies (e.g., \cite{Savage93}). The \ion{Fe}{2}
$\lambda 2344 $ and \ion{Fe}{2} $\lambda 2374$ are stronger than
expected based on the transition strengths relative to the other
lines, and \ion{Fe}{2} $\lambda 2374$ is redward by $\simeq 2$~\AA\
from the expected wavelength.  These lines probably suffer line
contamination, most likely from the \lya\ forest.  In the other cases,
absorption from the \ion{Fe}{2} transitions have equivalent widths and
ratios typical compared to \ion{Mg}{2} for absorption in the Galactic
halo.  However, the resolution is inadequate to allow for further
improvement over previously published results.

\section{Discussion of Individual Absorption Systems}

\subsection{$z \sim 1.166$}

Absorption lines identified at $\za \sim 1.166$ are \ion{O}{6}
$\lambda\lambda 1031$, 1037, \ion{Si}{3} $\lambda 1206$, \ion{Si}{4}
$\lambda 1393$ and \lya.  Although no other absorption lines are
present in this system at the detection threshold, \ion{N}{5} $\lambda
1238$ is coincident with the large absorption feature of \lya\ ($\za
\sim 1.207$) observed at $\lambda = 2683$~\AA.  Evidence for
absorption from the weaker doublet line, \ion{N}{5} $\lambda 1242$,
may be implied by the complex absorption feature centered around
$\lambda = 2693$~\AA\ (line~31 in Table~\ref{assoc_table}).  This
wavelength region is in a knotty section of the \lya+\ion{N}{5}
emission line which does not allow for useful measurements of the
column density.  If we place an upper limit on the equivalent width of
\ion{N}{5} at the detection threshold, the column density is
$N($\ion{N}{5}$) \lesssim 10^{14.0}$~\cmsq.  The detection of
\ion{Si}{4} $\lambda 1393$ is weak, $SL =2.8$.  Since this system
contains strong \ion{O}{6} and \ion{Si}{3}, \ion{Si}{4} is expected
and we consider this line identification real for the remainder of the
discussion.

Since the relative velocity between the QSO emission lines and this
absorption system is $\Delta v \simeq -5000$~\kms,  we classify this
system as associated to \qso.  The detection of \ion{O}{6} indicates a
high-ionization level which is usually assumed to indicate that the
absorber is ionized by the QSO, a not uncommon feature of $\za \sim \ze$
systems (\cite{Savaglio97}; \cite{Hamann97d}; \cite{Savage98}).
However, \ion{O}{6} absorption is also known (in at least one case) to
arise in a low-density halo ionized by the UV-background (Savage et
al.\ 1998).

A consistent model for the absorption must account for the presence of
\ion{Si}{3}, \ion{Si}{4}, \ion{O}{6}, and lack of \ion{N}{5}.  As
discussed in Hamann (1997), the detection of \ion{O}{6} and absence of
singly ionized metals (such at \ion{C}{2} and \ion{Si}{2}) in a
single-zone model requires an ionization parameter in the range, $U
\approx 0.01-1.0$, where $U \equiv n_{\gamma}/n_e$ is the
dimensionless ratio of the density of hydrogen-ionizing photons to the
electron density.  This range of ionization is expected to produce a
substantial fraction of \ion{N}{5} relative to other ions, e.g., $U
\sim 0.01-1.0$ is expected to produce $N($\ion{N}{5}$)/N($\ion{O}{6}$)
\sim 0.1-10$.  Therefore, its absence in the
single-zone model indicates the nitrogen abundance in the gas must be
depleted relative to oxygen and silicon.  This implies the absorption
may arise in a low metallicity environment which agrees with models of
galactic chemical evolution since both oxygen and silicon are produced
at early epochs in metal-poor, Population~II stars (so called
$\alpha$-elements) while nitrogen is dominantly produced in successive
generations of Population~I stars via the CNO cycle.  

Instead of a single value for the ionization parameter, an alternative
scenario is that a range of values exists.  The next simplest case is
a two-zone model where a high value for the ionization parameter, $U
\sim 1$, is responsible for the \ion{O}{6} absorption, while the
\ion{Si}{3} and \ion{Si}{4} originate in a region with $U \sim 0.003$.
If we compare the two-zone model to a galactic-like structure, the low
ionization lines arise in a disk-like, high-density region ($n_e \sim
1\; {\rm cm}^{-3}$), while the \ion{O}{6} absorption originates in a
high-temperature, low-density ($n_e \lesssim 0.1\; {\rm cm}^{-3}$),
pressure confining halo which corotates with the disk .

The origin of the absorption in this system remains uncertain.  It may
originate in a neighboring structure, separated in velocity by the
cosmological expansion or peculiar motions along the line of sight.
An alternative possibility is the absorption arises in a substantial
outflow from the QSO engine.  However, the latter scenario may be
inconsistent with other $\za \sim \ze$ systems since they typically
have {\em enhanced} \ion{N}{5} with respect to \ion{O}{6} (cf.\
\cite{Hamann97d}).  The basic arguments show that this system is
consistent with the absorption originating in either a metal depleted,
low-density halo, or a line of sight passing through multiple zones of
a neighboring galaxy.  Therefore, the origin remains unclear, and
there is insufficient evidence to classify the system as intrinsic to
\qso.

\subsection{$z \sim 1.199$}

We detect \lya\ and \lyb\ at $\za \sim 1.199$ with rest equivalent
widths given in Table~\ref{assoc_table}.  The Doppler parameter of
\lyb\ is broad, but the observed wavelength of this line is in the
vicinity of strong \ion{O}{6} lines from other systems and resides in
the low S/N portion of the \fos\ detector.  The residuals from the
best fit to the line profile shows an underlying complex structure.
The minimum neutral hydrogen column density is $N($\ion{H}{1}$)
\gtrsim 10^{14.6}$~\cmsq.  No metal absorption lines are present in
this system and the strongest lines expected have equivalent width
upper limits at the detection threshold, $W_{\rm obs} \lesssim
0.25$~\AA.  This is in stark contrast to the metal absorption lines in
the $z\:=\:1.166$ system where $N$(H~I) is almost identical.  The
corresponding upper limits on the strongest metal lines are
$N($\ion{Si}{2}$) \lesssim 10^{12.8}$~\cmsq, $N($\ion{O}{1}$) \lesssim
10^{14.1}$~\cmsq, $N($\ion{O}{6}$) \lesssim 10^{13.9}$~\cmsq, and
$N($\ion{N}{5}$) \lesssim 10^{14.0}$~\cmsq.  The absorber is
likely metal poor.  Since this absorber has a relative velocity of
$\Delta v \simeq 400$ km s$^{-1}$, it is under the influence of the
QSO.  However, without stronger knowledge of other lines in this
system, it is not possible to determine the ionizing source, e.g., the
QSO or the UV background. Evidence for the origin of the system is
inconclusive given only the detection of two Lyman series lines. The
existence of a metal poor absorber so close in redshift to a high
metallicity one remains intriguing, however.

\subsection{$z \sim 1.207$}

The absorption system of $z \sim 1.207$ is highly ionized given the
presence of \ion{O}{6} $\lambda\lambda 1031$, 1037, \ion{Si}{3}
$\lambda 1206$, \ion{N}{5} $\lambda\lambda 1238$, 1242, \ion{Si}{4}
$\lambda 1393$, 1402, \lya\ and \lyb, and the lack of any
low-ionization states including \ion{C}{1} $\lambda 1277$, \ion{C}{2}
$\lambda 1334$, \ion{N}{1} $\lambda 1200$, \ion{N}{2} $\lambda 1083$,
\ion{O}{1} $\lambda 1302$, and \ion{Si}{2} $\lambda 1260$. The
expected wavelength of the \ion{C}{4} $\lambda\lambda 1548$, 1550
doublet is $\sim 3420$~\AA\ and is redward of the spectral range of
the \fos\ G270H grating.  The absorption system has large equivalent
widths and the doublets are well resolved.  The properties of these
lines are listed in Table~\ref{assoc_table}, and the column densities
are listed in Table~\ref{column_depth_table}.  The overall relative
velocity of the absorption system is redward of the emission redshift
of \qso\ by $\Delta v \simeq 680$ km s$^{-1}$ implying infall toward
the emission source.

We observe an absorption feature at $\lambda \simeq 2467$~\AA\ which
the line detection software assigns a significance level of $SL =
2.8$.  Using a Gaussian fit, {\tt SPECFIT} calculates a larger value,
$W_{\rm FIT}/\sigma(W_{\rm FIT}) \simeq 3.4$, and a total equivalent
width, $W \simeq 0.5$~\AA.  Since this line resides in the \lya\
forest, our original identification was intervening \lya.  The
observed wavelength of this line also corresponds to \ion{P}{5}
$\lambda 1117$ at $\za = 1.2066$.  Since the lines in this system are
strong, the detection of \ion{P}{5} is not necessarily unexpected.
This atomic transition occurs as a doublet, and the other doublet
member, \ion{P}{5} $\lambda 1128$, is redshifted to $\simeq 2490$~\AA.
Based on the oscillator strengths, the \ion{P}{5} $\lambda 1128$ line
is expected to be weaker by up to a factor of two compared to
\ion{P}{5} $\lambda 1117$. There is a strong absorption line at
$\lambda = 2489$~\AA\ with $SL \simeq 7$ that exhibits a complex
profile.  If the \ion{P}{5} $\lambda 1117$ identification of this line
is correct, then the absorption at $\simeq 2490$~\AA\ must result from
blending of multiple lines (probably from the \lya\ forest) which are
unresolved at the resolution of \fos.  Since the detection of
\ion{P}{5} has potentially important implications for the metal
enrichment and ionization states of intrinsic absorption systems (cf.\
\cite{Hamann98}), we consider this line a {\em possible}
identification of \ion{P}{5} $\lambda 1117$.

The profiles for the nine lines in this system are shown in Figure~\ref{line_profiles} as normalized flux versus velocity relative to the
nominal redshift value, $\za = 1.207$.  The absorption lines may
consist of several, narrow components with slightly varying (and
indeterminable) relative velocities and covering factors.  When
observed with the modest seeing of the FOS, they appear as a single,
broader (FWHM $\sim 200$~\kms) line.  This system is a candidate for
intrinsic absorption since it is close to the QSO emission in velocity
space and is highly ionized.

\subsubsection{Ionization}

The dimensionless ionization parameter, $U \equiv n_\gamma / n_e$, is
a useful parameterization for the ionization state of the
absorbing gas.  Given the presence of highly ionized, strong
\ion{N}{5} and \ion{O}{6} and the absence of \ion{Si}{2} and
\ion{C}{2}, the ionization parameter is in the range
$0.01 \lesssim U \lesssim 1.0$.

We can investigate the ionization state of the system by rewriting the
ionization parameter as
\begin{equation}
	U = \frac{n_{\gamma}}{n_e} = \frac{1}{n_ec} \frac{d_L^2}{r^2}
	\int_{\nu_{\rm th}}^{\infty} \frac{\pi F_{\nu}}{h\nu} d\nu
\end{equation}
where $c$ is the speed of light, $h$ is the Planck constant, $n_e$ is
the electron density of the gas, $d_L$ is the luminosity distance from
the observer to the ionizing source (QSO engine), and $r$ is the
distance between the QSO and the absorbing material.  The hydrogen
ionizing threshold frequency is $\nu_{\rm th}$, and $F_{\nu}$ is the
observed spectral flux (erg~\cmsq~s$^{-1}$ Hz$^{-1}$).  To evaluate
the integral, we adopt a spectral energy distribution for the QSO
emission source that is similar to a typical radio-quiet AGN (cf.\
\cite{Hamann97}),
\begin{equation}
	F_\nu(\nu) \propto \nu^{\alpha_{{\rm uv}}} e^{-h\nu/kT_{\rm uv}} e^{-kT_{\rm ir}/h\nu} + \kappa(\nu) \nu^{\alpha_x} .
\end{equation}
The value of $T_{\rm uv}$ is chosen to reproduce the ultraviolet peak
typically seen in AGN spectra, and $kT_{\rm ir} = 0.136$ eV is an infrared
cut-off to prevent free-free heating at low energies.  The function
$\kappa(\nu)$ is chosen to produce to the correct relative strengths of the UV
and X-ray fluxes and incorporates cut-offs in the X-ray producing mechanism at
high and low energies.

To emulate the spectral energy distribution of \qso, we adopt $T_{\rm
uv} = 10^6$ K and spectral indices of $\alpha_{\rm uv} = -1.1$,
$\alpha_x = -1.1$ which empirically agrees with the FOS and ROSAT
observations (\S 2; \cite{Yuan98}).  From this, we extrapolate a
rest-frame optical flux density of $F_{\nu}($2500~\AA$) \simeq 2.9
\times 10^{-27}$ erg~\cmsq~s$^{-1}$ Hz$^{-1}$ and derive an X-ray to
optical two-point power law index, $\alpha_{ox} = -1.9$ [$\alpha_{ox}
\equiv 0.384 \log(F_\nu($2~keV$)/F_\nu($2500~\AA$))$].  Using equation
(3) and a fiducial value for the deceleration parameter, $d_L(q_0 =
0.1) = 5348h^{-1}$~Mpc, equation (2) can be evaluated and rewritten as
\begin{equation}
	r\; n_6^{1/2} = 11\; U^{-1/2} h^{-1}\; {\rm pc},
\end{equation}
where $n_6$ is the electron density in units of $10^6$~cm$^{-3}$, and
here, $h$ is the dimensionless Hubble parameter, $H_0 = 100
h$~\kms~Mpc$^{-1}$.

We investigate the ionization quantitatively with the photoionization
code, {\tt CLOUDY} (v. 90.03, \cite{Ferland96}).  For the ionizing
source, we use the spectrum defined above.  The absorber is modeled as
a plane-parallel slab, illuminated on one side by the QSO.  We tested 
a range of absorber electron densities, $n_e = 1 - 10^{10}$~cm$^{-3}$, 
and observed that the ionization level is
approximately constant for a given value of $U$ in the range of
interest ($0.01 \lesssim U \lesssim 1.0$).  Therefore, we used a fixed
electron density, $n_e = 10^6$~cm$^{-3}$, but note in passing that
changes in the electron density do not significantly alter the
qualitative results. The metallicity of the gas is normalized to solar
values from a compilation of meteoritic, photospheric and coronal
abundances (\cite{Grevesse89}).  To examine the state of the absorbing
medium for a range of input parameters, we used {\tt CLOUDY} to compute
the output column densities, ionization fractions, and electron
temperatures at a given ionization parameter for gas of specified
total hydrogen column density and metallicity.  Caution must be used
when interpreting the results of the photoionization code since it
assumes the absorber is in photoionizational equilibrium.  This is not
established {\em a priori} and may produce misleading results (cf.\
\cite{Krolik95}).

\begin{figure*}
\figurenum{4}
\epsscale{1.2}
\plotone{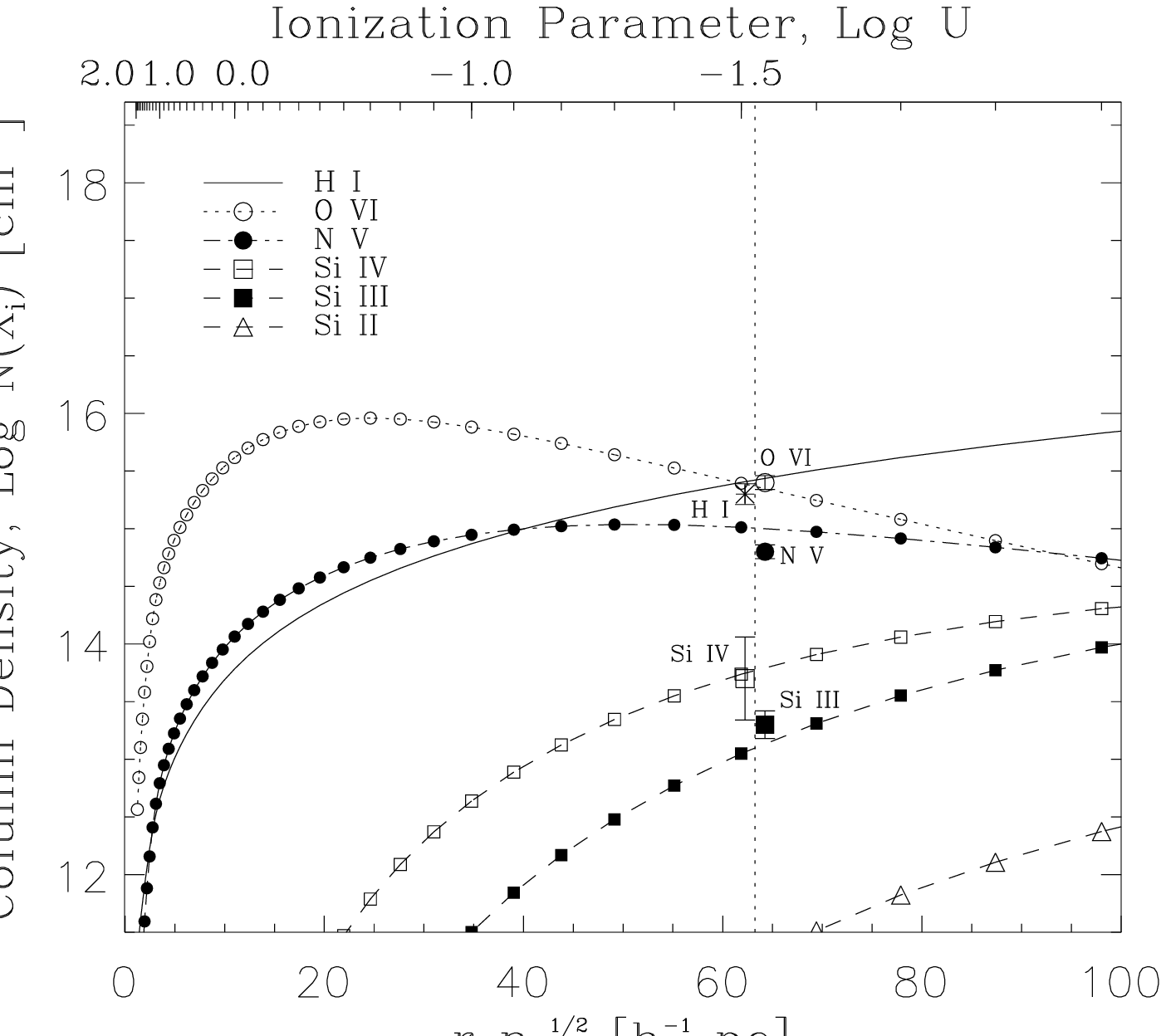}
\caption{The computed  column densities from {\tt CLOUDY} are shown as a function of the value of $r\; n_6^{1/2}$ from the model for the  $\za \sim 1.207$ absorption system.  The values of the ionization parameter are shown along the top abscissa in logarithmic units. The measured  column densities are shown as points with $3\sigma$ error bars and are staggered around the nominal value, $r\, n_6^{1/2} = 63\; h^{-1}$~pc. The absorber model has a total hydrogen column density, $\NH \simeq 2\times 10^{19}$~\cmsq, a metallicity, $Z \simeq 2 Z_\odot$, and constant electron density.  \label{cloudy_results}}
\end{figure*}

The column densities which were deduced in \S3.2 provide constraints
on the permitted conditions of the absorbing material.  Assuming these
values are a good approximation of the true column densities (i.e.\ no
substantial, hidden columns exist), the conditions of the absorber are
almost uniquely reproducible (aside from systematic effects).  We show
in Figure~\ref{cloudy_results} the {\tt CLOUDY} results for the
single--zone absorption model with a total hydrogen column density,
$\NH \approx 2\times 10^{19}$~\cmsq, and metallicity, $Z \sim 2
Z_\odot$.  The column densities calculated by the model are shown
versus a range of ionization parameters, $U$.    Based on
the relative velocity--ionization energy relation which we observe in
this absorption system (see \S 4.3.2), it is unlikely that a
single-zone model is a completely adequate description.  The results
presented here are, therefore, a simple paradigm for the ionization in
this absorber.  The model
simultaneously reproduces the observed column densities for $\log U
\simeq -1.5$ and corresponding electron temperature, $T_e \sim 2\times 10^4$~K.

The derived ionization parameter corresponds to spatial dimensions of
$r\; n_6^{1/2} \simeq 63\; h^{-1}$~pc.  However, without a constraint
on the electron density of the absorber, the distance to the ionizing
source remains unknown.  The presence of \ion{O}{6} indicates the
lines are highly ionized and is evidence for intrinsic absorption.  
The supersolar metallicity which is needed to model the column densities in this system is also common in intrinsic absorption systems. However, as discussed in \S4.1, these features are not exclusive to intrinsic absorbers.   An investigation of the time-variability nature
of the absorption lines could provide a useful upper limit on the
recombination time, ($t_{\rm rec} \sim 1/\,n_e\alpha_r$, where
$\alpha_r$ is the recombination coefficient), which in turn would
establish a lower limit on the electron density and thus a constraint
on the distance from the ionizing source.  The ionization state of the
system alone does not distinguish the location of the absorber.

Since the derived column densities only provide constraining lower
limits, the model depicted in Figure~\ref{cloudy_results} is not
unique.  ROSAT X-ray observations provide some insight into the total
absorbing hydrogen column density lying along the sight line, $\NH =
4.6^{+3.9}_{-2.9} \times 10^{20}$~\cmsq\ (\cite{Yuan98}). However, the
contribution of neutral hydrogen from the Milky Way is estimated to be
$N($\ion{H}{1}$)\simeq3.3\times10^{20}$~\cmsq, meaning that any
further absorption would come from a column of $\sim\:1\times
10^{20}$~\cmsq. Our adopted value of $\NH \simeq 2\times
10^{19}$~\cmsq\ is therefore not inconsistent with the X-ray data.
Since our derived column densities are only estimates due to the
nature of the FOS observations, there is a degeneracy between the
total column density and the metallicity in the absorption models.
Using our results from {\tt CLOUDY}, for $\NH \lesssim 10^{19}$ \cmsq,
the metallicity must be significantly higher to account for the
observed columns.  Subsolar metallicities are likewise not completely
excluded, but substantially more $N($\ion{H}{1}$)$ is expected
relative to the metallic columns derived from the FOS data in this
case.

\begin{figure*}
\figurenum{5}
\epsscale{1.33333}
\plotone{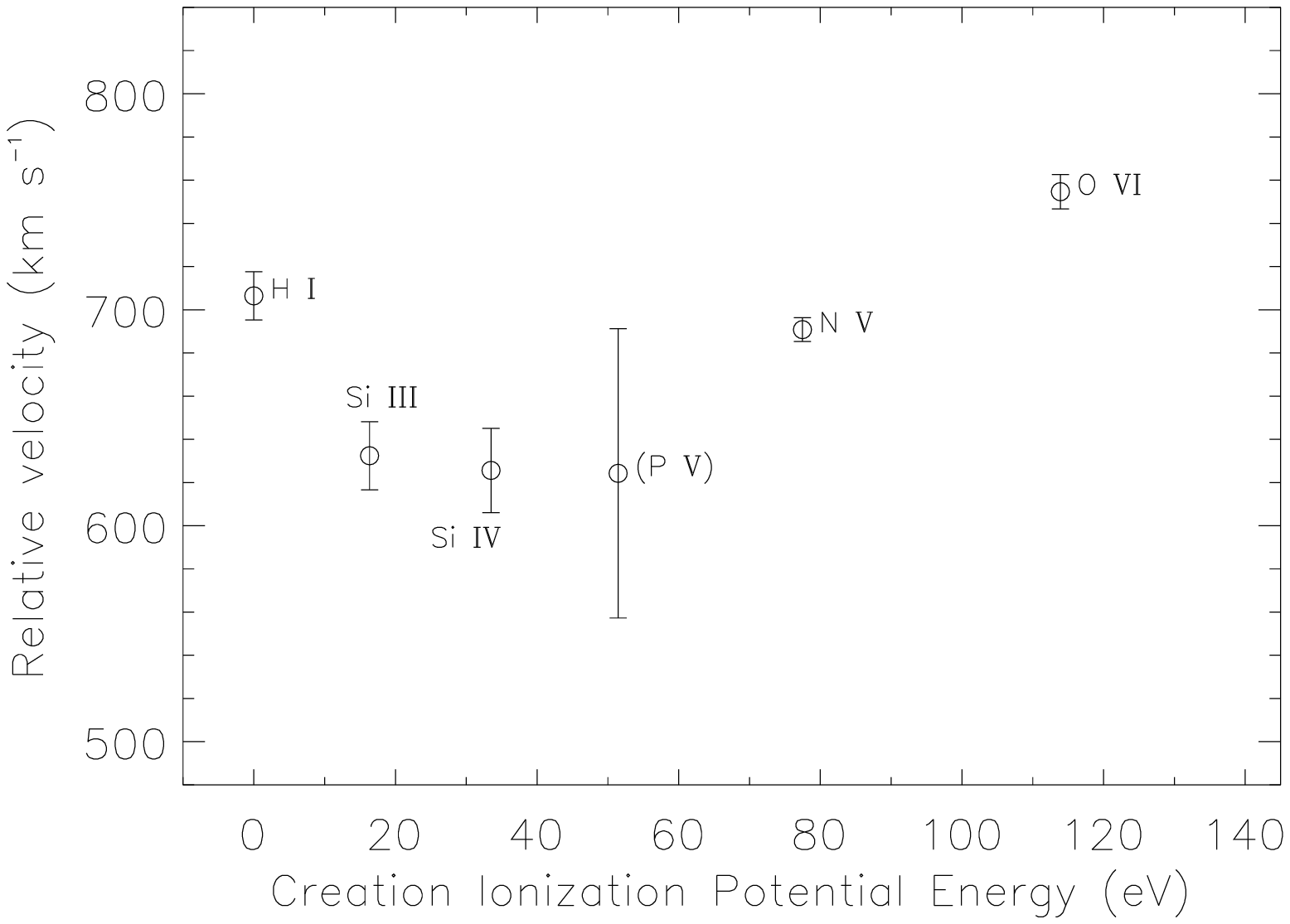}
\caption{The velocity for each absorption line in the $\za \sim 1.207$ system relative to the $\ze = 1.202$ emission redshift is plotted versus the creation energy required for each ion.  The redshifts are taken from Table~\ref{assoc_table}.  The error bars represent the propagation of the $1 \sigma$ uncertainties from the line fitting analysis (see \S 2).  Infall is implied by the direction of the relative velocity.  \ion{P}{5} is included to show its relationship to the other lines in this system.\label{IonGrad} }
\end{figure*}

\subsubsection{Relative Velocity of the Gas Flow and Multi-Component Absorption}

The positive relative velocity for the $\za \sim 1.207$ absorption
system implicitly indicates the absorbing gas is infalling toward the
central AGN, yet the systemic redshift could be slightly larger than
that adopted for \qso, and outflow cannot be entirely excluded. Infall
has been observed in several other $\za \sim \ze$ metal systems
(Savaglio et al.\ 1997; \cite{Dobrzycki99}) and in several \lya\
absorption systems in a \lya\ emission line survey of radio galaxies
(\cite{Ojik97}).

The absorption lines in this system contain varying velocities
relative to the emission redshift ($600 \lesssim \Delta v \lesssim
800$~\kms) and span a wide range of ionization.  Furthermore, these
properties are correlated in the sense that the centroids of the more
highly excited lines have higher relative  velocities.  If the
absorption system contains multiple components, then the ionization
state depends on the density and distance from the ionizing source for
each component (in this case the QSO engine).

We quantify this relationship by comparing the relative velocity of
each line with the {\em creation ionization potential energy} (CIP)
required to produce the line.  The CIP of ion $X_i$ is defined as the
energy required to ionize the state $X_{i-1}$ to $X_i$.  The second
column of Table~\ref{column_depth_table} contains the CIP for each
ion.   In Figure~\ref{IonGrad}, we plot the velocity of each ion,
relative to the emission redshift, versus the CIP.

An explanation for this correlation is that the more highly ionized
absorption line components are shifted blueward of less ionized ones.
Clearly, the CIP is related to the ionization parameter
since at higher values of $U$, more ionizing photons are available to
produce the more highly excited ions. Using the relationship in
equation (4), the more blueward line components have lower values of
$r\; n_6^{1/2}$.  The simplest model has two absorption components,
with higher and lower levels of ionization, and a velocity separation
of $\sim 200$ \kms.  The convoluted sum of these components results in
a shift of the observed absorption line centroids.  The column densities of the lower
ionization species (e.g., \ion{Si}{3}, \ion{Si}{4}) result mostly from 
the less ionized component and the column densities with the highest ionization (e.g.,
\ion{O}{6}) result mainly from the higher ionized component with
little contribution from the other.

It is interesting that the relative velocity of the \ion{H}{1} lines
do not follow the correlation and are approximately equal to the value
of the \ion{N}{5} doublet.  In the context of a two-component
model, most of the \ion{H}{1} absorption occurs from the highly
ionized line component.  This is plausible since, unlike the other
elements in this system, there are no higher ionization states for
hydrogen.  It is also possible that if the more highly ionized component
occults the other and is not optically thin, then it may shield the lower ionization component from part of the ionizing continuum.  Therefore, there is not necessarily any
discrepancy between the observed correlation and the relative velocity
of \ion{H}{1}.

The correlation between relative velocity and ionization level has not
been reported in other systems.  We examined 17 other QSO absorption
line systems (both associated and intervening) from 11 QSO lines of
sight (PKS2135$-$147, \cite{Hamann97d}; H1821+643, Savage et al.\
1998; PKS0123+257, \cite{Barlow97}; Q000$-$2619, Savaglio et al.\
1997; Q2116$-$358, Wampler et al.\ 1993; Q0347$-$241, Q2116$-$358,
M{\o}ller et al.\ 1994; HS1946+7658, \cite{Tripp96}; Q2343+125, Hamann
et al.\ 1997c; PKS0424$-$131, Q0450$-$131, Petitjean et al.\ 1994),
and none of the intervening absorption systems bear a velocity
correlation with CIP.  Only one other associated
absorption system, Q0450$-$131 (Petitjean et al.\ 1994), may suggest a similar correlation.  In
this system, the relative velocity between the emission and absorption
system is $\Delta v \sim -2100$~\kms\ and there is a similar trend
between ionization strength and relative velocity.  However, the
ionization only spans the range \ion{Si}{3} to \ion{N}{5} (\ion{O}{6}
was blueward of the spectral grating) and any correlation among the
few identified lines is only suggestive at best.  Therefore, the $\za
\sim 1.207$ absorption system of \qso\ is possibly the first
observation of a CIP--velocity correlation in an associated QSO absorption
system. Why only this system should show such an effect is unclear.

The origin of the absorption system remains inconclusive.
High-resolution, UV spectroscopy may provide the answer.  If the
multiple absorbing components are resolved with the improved
resolution, this would permit a detailed analysis of the individual
constituents. If the line components are well resolved and separated
by the few hundred \kms\ then the origin of the system might be from
halos of neighboring galaxies moving in the potential of the QSO
cluster.  However, the presence of significant \ion{N}{5} (and
possible \ion{P}{5}) and high overall metallicity is unusual for
extended galactic features. Alternatively, the broad nature of the
absorption profiles may remain even when viewed at high resolution.
This feature, combined with the high level of ionization and high
metallicity, is similar to the BAL phenomenon and may suggest that the
absorber consists of many dense components with a small volume filling
factor.  The problem with this scenario is that broad, intrinsic lines
typically show {\em outflow} velocities of up to thousands of \kms\
compared to the emission redshift of the QSO.  This is in contrast the
system which is presented here.  It is also possible that both scenarios to
contribute to this system. Therefore, the source of this absorber is
not clearly evident.

\section{Summary}

We have analyzed the three associated ($\za \sim \ze$) absorption
systems identified in the spectrum of \qso\ over the wavelength range
$\sim 2220-3280$~\AA\ using \hst/\fos\ data.  All the systems have
relative velocities within $\sim 5000$ \kms\ of the emission redshift
and are highly ionized.

The associated system at $\za \sim 1.166$ shows highly ionized
\ion{O}{6}, \ion{Si}{3}, and weak \ion{Si}{4}.  The large
deconvolved FWHM ($\sim 450$~\kms) of the \ion{O}{6} lines indicate
there are probably unresolved narrow subcomponents with a velocity
dispersion along the line of sight of a few hundred~\kms.  Since this
system is separated from the emission redshift by $\Delta v \simeq
5000$~\kms\ and contains \ion{O}{6}, \ion{Si}{3}, and \ion{Si}{4} but
is devoid of other highly ionized lines to the limits of detection
[e.g., $N($\ion{N}{5}$) \lesssim 10^{14}$~\cmsq], its origin is
unclear.
	
Only \lya\ and \lyb\ are identified in the absorption system at $\za
\sim 1.199$, with a neutral hydrogen column density of
$N($\ion{H}{1}$) = 10^{14.6}$~\cmsq.  No metallic ions are detected in
this system and the upper limits on the strongest expected metal lines
are $N($\ion{Si}{2}$) \lesssim 10^{12.8}$~\cmsq, $N($\ion{O}{1}$)
\lesssim 10^{14.1}$~\cmsq, $N($\ion{O}{6}$) \lesssim 10^{13.9}$~\cmsq,
and $N($\ion{N}{5}$) \lesssim 10^{14.0}$~\cmsq.  The origin of this
system is inconclusive given only the presence of two Lyman-series
lines.

The most complex $\za \sim \ze$ system is at $\za \sim 1.207$ and
shows prominent absorption due to \lya, \lyb, \ion{N}{5}, \ion{O}{6},
\ion{Si}{3}, \ion{Si}{4}, and possibly, \ion{P}{5}.   Comparisons with
photoionization simulations show that the minimum column densities
derived for the absorption lines of this system are consistent with an
absorber of metallicity, $\sim 2Z_\odot$, and
total hydrogen column density, $\NH \simeq 2 \times 10^{19}$~\cmsq,
which is ionized by the incident radiation of \qso.  If the
identification of \ion{P}{5} $\lambda 1117$ in this associated system
is valid, then this suggests that the ionization and optical depth may be
very high.  The absorption lines have individual relative velocities
$\Delta v \simeq 600 - 800$~\kms\ toward the QSO (e.g., infall) and
these are correlated with the creation ionization potential energy for each
transition.

The origin of the absorber remains inconclusive.  The highly-ionized
species of \ion{O}{6} and \ion{N}{5} (and possibly \ion{P}{5}), with
broad absorption profiles and high overall metallicity are common
features of intrinsic absorption systems.  If the absorption lines
remain unresolved at higher resolution, then the relative
velocity--ionization energy relationship may indicate the absorber contains many
small, dense clumps with a low volume filling factor.  However, if the
absorption lines result from the convolution of several
components, then the data is equally well explained by
absorption components from nearby galaxies moving in the potential of
the QSO cluster environment.  Line-variability investigations
could yield constraints on the absorber density and distance from the
ionizing source which could definitively yield the location of the
absorber.  Therefore, the origin of this absorption system will
undoubtedly remain uncertain until future observations of \qso\
become available.

\acknowledgements

We thank our colleagues at the Johns Hopkins University and the Space
Telescope Science Institute for stimulating discussions on this
subject, particularly R. Wyse, E. Agol, C. Pickens, and A. Zirm.  We
acknowledge the {\em Hot Topics in Astrophysics} group at JHU for
valuable comments.  Special thanks to the anonymous referee whose
suggestions greatly improved the quality of this paper.  Also, thanks
to D. Schneider for supplying the QALKP line search software.  Support
for this work was provided by NASA through grants GO-03676.01,
GO-02644.01, GO-06728.01 and GO-06707.01 from the Space Telescope
Science Institute, which is operated by the Association of
Universities for Research in Astronomy, Inc., under NASA contract NAS
5-26555.


\begin{thebibliography}{999}
\bibitem[Aldcroft, Bechtold, \& Foltz 1997]{Aldcroft97} Aldcroft, T.,
Bechtold, J., \& Foltz, C. 1997, in ASP Conf. Prov. 128, Mass Ejection
from AGN, ed.\ R. Weymann, I. Shlosman, \& N. Arav (San Fransisco:
ASP), 25
\bibitem[Anderson et al.\ 1987]{Anderson87} Anderson, S. F., Weymann,
R. J., Foltz, C. B., \& Chaffee, F. H., Jr. 1987, \aj, 94, 278
\bibitem[Arp \& Duhalde 1985]{Arp85} Arp, H., \& Duhalde, O. 1985,
\pasp, 97, 1149
\bibitem[Bahcall et al.\ 1993]{Bahcall93} Bahcall, J. N., et al.\
1993, \apjs, 87, 1
\bibitem[Barlow et al.\ 1992]{Barlow92} Barlow, T. A., Junkkarinen,
V. T., Burbidge, E. M., Weynmann, R. J., Morris, S. L., \& Korista,
K. T. 1992, \apj, 397, 81
\bibitem[Barlow \& Sargent 1997]{Barlow97} Barlow, T. A., \& Sargent,
W. L. W. 1997 \aj, 113, 136
\bibitem[Barthel et al.\ 1988]{Barthel88} Barthel, P. D., Miley,
G. K., Schilizzi, R. T., \& Lonsdale, C. J. 1988,~\aap, 73, 515
\bibitem[Barthel, Tytler, \& Thomson 1990]{Barthel90} Barthel, P. D.,
Tytler, D. R., \& Thomson, B. 1990,~\aap, 82 339
\bibitem[Bowen, Blades, \& Pettini 1995]{Bowen95} Bowen, D. V.,
Blades, J. C., \& Pettini, M. 1995, \apj, 448, 634
\bibitem[Dobrzycki, Engels, \& Hagen 1999]{Dobrzycki99} Dobrzycki, A.,
Engels, D., \& Hagen, H.-J. 1999, \aap, 349, L29
\bibitem[Espey 1993]{Espey93} Espey, B. R. 1993, \apjl, 411, L59
\bibitem[Evans 1993]{Evans93} Evans, I. N. 1993, \fos\ Instrument
Science Rep. (CAL/\fos-104) (Baltimore: STScI)
\bibitem[Ferland 1996]{Ferland96} Ferland, G. J. 1996, ``HAZY: a Brief
Introduction to Cloudy,'' Univ. Kentucky, Dept. Physics \& Astron.,
Internal Rep.
\bibitem[Ferland et al.\ 1996]{Ferland96b} Ferland, G. J., Baldwin,
J. A., Korista, K. T., Hamann, F., Carswell, R. F., Phillips, M.,
Wilkes, B., \& Williams, R. E. 1996, \apj, 461, 683
\bibitem[Foltz et al.\ 1988]{Foltz88} Foltz, C. B., Chaffee, F. H.,
Jr., Weymann, R. J., \& Anderson, S. F. 1988, in QSO Absorption Lines:
Probing the Universe, ed.\ J. C. Blades, D. A. Turnshek, \&
C. A. Norman (Cambridge: Cambridge Univ. Press), 53
\bibitem[Foltz et al.\ 1986]{Foltz86} Foltz, C. B., Weymann, R. J.,
Peterson, B. M., Sun, L., Malkan, M. A., \& Chaffee, F. H., Jr. 1986,
\apj, 307, 504
\bibitem[Grevesse \& Anders 1989]{Grevesse89} Grevesse, N., \& Anders,
E. 1989, in AIP Conf Proc. 183, Cosmic Abundances of Matter, ed.\
C.I. Waddington (New York: AIP), 1
\bibitem[Hamann 1997]{Hamann97} Hamann, F. 1997, \apjs, 109, 279
\bibitem[Hamann 1998]{Hamann98} Hamann, F. 1998, \apj, 500, 798
\bibitem[Hamann et al.\ 1995]{Hamann95} Hamann, F., Barlow, T. A.,
Beaver, E. A., Burbidge, E. M., Cohen, R. D., Junkkarinen, V., \&
Lyons, R. 1995, \apj, 443, 606
\bibitem[Hamann et al.\ 1997a]{Hamann97a} Hamann, F., Barlow, T. A.,
Cohen, R. D., Junkkarinen, V. T., \& Burbidge, E. M. 1997a, in ASP
Conf. Proc. 128, Mass Ejection From AGN, ed.\ R. Weymann, I. Shlosman,
\& N. Arav (San Fransisco: ASP), 19
\bibitem[Hamann et al.\ 1997b]{Hamann97b} Hamann, F., Barlow, T. A.,
Cohen, R. D., Junkkarinen, V. T., \& Burbidge, E. M. 1997b, in ASP
Conf. Proc. 128, Mass Ejection From AGN, ed.\ R. Weymann, I. Shlosman,
\& N. Arav (San Fransisco: ASP), 187
\bibitem[Hamann, Barlow, \& Junkkarinen 1997c]{Hamann97c} Hamann, F.,
Barlow, T. A., Junkkarinen, V. T. 1997c, \apj, 478, 87
\bibitem[Hamann et al.\ 1997d]{Hamann97d} Hamann, F., Beaver, E. A.,
Cohen, R. D., Junkkarinen, V., Lyons, R. W., \& Burbidge, E. M. 1997d,
\apj, 488, 155
\bibitem[Hamann \& Ferland 1993]{Hamann93} Hamann, F., \& Ferland,
G. 1993, \apj, 418, 11
\bibitem[Hamann \& Ferland 1999]{Hamann99} Hamann, F., \& Ferland,
G. 1999, \araa, in press ({\tt astro-ph/9904223})
\bibitem[Krolik \& Kriss 1995]{Krolik95} Krolik J. H., \& Kriss,
G. A. 1995, \apj, 447, 512
\bibitem[Kriss 1994]{Kriss94} Kriss, G. A. 1994, in
A.S.P. Conf. Series, Vol. 61, Astronomical Data Analysis Software \&
Systems III, ed.\ D. R. Crabtree, R. J. Hanisch, \& J. Barnes (San
Francisco: ASP), 437
\bibitem[Matteucci \& Padovani 1993]{Matteucci93} Matteucci, F, \&
Padovani, P. 1993, \apj, 419, 485
\bibitem[M{\o}ller, Jakobsen, \& Perryman 1994]{Moller94} M{\o}ller,
P., Jakobsen, P., \& Perryman, M. A. C. 1994,~\aap, 287, 719
\bibitem[Norman et al.\ 1996]{Norman96} Norman, C. A., Bowen, D. V.,
Heckman, T., Blades, J. C., \& Danly, L. 1996, \apj, 472, 73
\bibitem[Petitjean, Rauch, \& Carswell 1994]{Petitjean94} Petitjean,
P., Rauch, M., \& Carswell, R. F. 1994,~\aap, 281, 331
\bibitem[Savage et al.\ 1993]{Savage93} Savage, B. D., et al.\ 1993,
\apj, 413, 116
\bibitem[Savage, Tripp, \& Lu 1998]{Savage98} Savage, B. D., Tripp,
T. M., \& Lu, L. 1998, \apj, 115, 436
\bibitem[Savaglio et al.\ 1997]{Savaglio97} Savaglio,
S.,  Cristiani, S., D'Odorico, S., Fontana, A., Giallongo, E., \& Molaro, P. 1997, \aap, 318, 347
\bibitem[Smith \& Heckman 1990]{Smith90} Smith, E. P. \& Heckman,
T. M. 1990, \apj, 348, 38
\bibitem[Schneider et al.\ 1993]{Schneider93} Schneider, D., et al.\
1993, \apjs, 87, 45
\bibitem[Tripp, Limin, \& Savage 1996]{Tripp96} Tripp, T. M., Limin,
L., \& Savage, B. D. 1996, \apjs, 102, 239
\bibitem[Turnshek et al.\ 1996]{Turnshek96} Turnshek, D. A., Kopko,
M., Monier, E., Noll, D. Espey, B., \& Weymann, R. J. 1996, \apj, 463,
110
\bibitem[Tytler \& Fan 1992]{Tytler92} Tytler, D., \& Fan, X.-M.,
1992, \apjs, 79, 1
\bibitem[van Ojik et al.\ 1997]{Ojik97} van Ojik, R., R\"{o}ttgering,
H. J. A., Miley, G. K., \& Hunstead, R. W. 1997, \aap, 317, 358
\bibitem[Verner, Barthel, \& Tytler 1994]{Verner94} Verner, D. A.,
Barthel, P. D., \& Tytler, D., 1994,~\aaps, 108, 287
\bibitem[V\'{e}ron-Cetty, \& V\'{e}ron 1998]{Veron98} V\'{e}ron-Cetty,
M. -P., \& V\'{e}ron, P. 1998, A Catalogue of Quasars and Active
Nuclei, 8th ed.\ to appear in ESO Scientific Report No. 18
\bibitem[Wampler, Bergeron, \& Petitjean 1993]{Wampler93} Wampler, E.,
J., Bergeron, J., \& Petitjean, P., 1993,~\aap, 272, 15
\bibitem[Weymann et al.\ 1998]{Weymann98} Weymann, R. J., et al.,
1998, \apj, 506, 1
\bibitem[Weymann et al.\ 1979]{Weymann79} Weymann, R. J., Williams
R. E., Peterson B. M., Turnshek, D. A. 1979, \apj, 234, 33
\bibitem[Yuan et al.\ 1998]{Yuan98} Yuan, W., Brinkmann, W., Siebert,
J., \& Voges, W. 1998,~\aap, 330, 108
\end{thebibliography}
\end{document}